\documentclass[onecolumn,12pt, floatfix, preprintnumbers, eqsecnum, letterpaper, superscriptaddress,
nofootinbib]{revtex4}
\usepackage{graphicx}
\usepackage{microtype}
\usepackage{amsmath}
\usepackage{amssymb}
\usepackage{subfigure}
\usepackage{hyperref}
\usepackage{url}
\usepackage{xcolor}
\usepackage{color}
\usepackage{mathrsfs}
\usepackage{calrsfs}
\usepackage{amsfonts}
\usepackage{eufrak}
\usepackage{tabularx}
\usepackage{eucal}
\usepackage{latexsym}
\usepackage{ragged2e}
\usepackage{epsfig}
\usepackage{textcomp}
\usepackage{caption}
\usepackage{subfigure}
\usepackage{amsmath}
\usepackage{graphicx}
\DeclareCaptionJustification{justified}{\leftskip=0pt \rightskip=0pt \parfillskip=0pt plus 1fil}
\definecolor{vividviolet}{rgb}{0.62, 0.0, 1.0}
\definecolor{amaranth}{rgb}{0.9, 0.17, 0.31}
\definecolor{palatinateblue}{rgb}{0.15, 0.23, 0.89}
\definecolor{brightpink}{rgb}{1.0, 0.0, 0.5}
\definecolor{cornflowerblue}{rgb}{0.39, 0.58, 0.93}
\definecolor{deepcarminepink}{rgb}{0.94, 0.19, 0.22}
\definecolor{radicalred}{rgb}{1.0, 0.21, 0.37}

\hypersetup{ linktoc=all,
    colorlinks, linkcolor={palatinateblue},
    citecolor={brightpink}, urlcolor={amaranth}
}

\graphicspath{{Images/}}

\graphicspath{{Images/}}

\def\sideremark#1{\ifvmode\leavevmode\fi\vadjust{\vbox to0pt{\vss% the remark
 \hbox to 0pt{\hskip\hsize\hskip1em%                          will appear only
 \vbox{\hsize1.3cm\tiny\raggedright\pretolerance10000%          on the side
 \noindent #1\hfill}\hss}\vbox to8pt{\vfil}\vss}}}%

\def\beq{\begin{equation}}
\def\eeq{\end{equation}}

\begin{document}

\title{Imprints of Topological Thermodynamics on Black Hole Dynamics
%Imprints of Topological Thermodynamics in the Dynamics of Black Hole Quasinormal Modes
}

\author{Yue Chu}
\email{ychu77@nuaa.edu.cn}
\affiliation{College of Physics, Nanjing University of Aeronautics and Astronautics, Nanjing, 211106, China}

%\author{Shi-Bei Kong}\email{shibeikong@ecut.edu.cn}\affiliation{School of Science, East China University of Technology, Nanchang 330013, Jiangxi, China}

\author{Chen-Hao Wu}
\email{chenhao\_wu@nuaa.edu.cn}
\affiliation{College of Physics, Nanjing University of Aeronautics and Astronautics, Nanjing, 211106, China}

%\author{Hongsheng Zhang}
%\affiliation{School of Physics and Technology, University of Jinan,336 West Road of Nan Xinzhuang, Jinan, Shandong 250022, China}

\author{Ya-Peng Hu\textsuperscript{*}}
\email{huyp@nuaa.edu.cn}
\affiliation{College of Physics, Nanjing University of Aeronautics and Astronautics, Nanjing, 211106, China}
\affiliation{Key Laboratory of Aerospace Information Materials and Physics (NUAA), MIIT, Nanjing 211106, China}

\begin{abstract}
By employing Duan's topological method, we can classify critical points based on their topological charge as \( Q = \pm1,0 \).
Previous work [Wei et al., PRD
 105, 104003 (2022)] investigate two typical anti-de Sitter (AdS) black hole: Reissner-Nordström (RN) case (with only one critical point \( Q = -1 \)) and Born–Infeld (BI) case (with two critical points \( Q = \pm 1 \)).
In this work, we first find that all three types of critical points exhibit in quantum anomalous black holes for 4-D spacetime.
We then compute the quasinormal modes of massless scalar perturbations near these critical points, and find that both oscillation frequency and damping rate increase with black hole radii at the critical temperature. 
Besides such common behavior, though the \( Q = +1,0 \) cases do not have a discernible pattern due to few samples, the \( Q = -1 \) case shows very similar dynamical characteristics across all three black hole solutions, which implies a non-trivial connection between topological thermodynamics and dynamics.

\end{abstract}

\maketitle

\section{Introduction}
Black hole thermodynamics in AdS spacetime exhibits rich thermodynamic behaviors and is believed to have multiple AdS/CFT correspondences \cite{Natsuume:2014sfa,Harlow:2014yka}, providing a deep connection between gravity and quantum field theory.
The relevant work starts from the discovery of the Hawking-Page transition \cite{Hawking:1982dh}, a thermal phase transition between Schwarzschild-AdS black holes and pure AdS spacetime driven by temperature variations.
Subsequent work revealed that charged AdS black holes undergo a first$-$order transition between small and large black hole phases, closely resembling the liquid$-$gas transition in a van der Waals fluid.
Dolan \cite{Dolan:2011xt} was the first to map the $(P,V)$ isotherms for charged AdS black holes, observing an analogy with the van der Waals $P-V$ diagram, and to determine its critical point.
A significant advancement came with Kubizňák and Mann's proposal to treat the cosmological constant as a thermodynamic pressure $P=-\Lambda/8\pi$ \cite{Kubiznak:2012wp}.
This identification enabled them to derive a van der Waals-type equation of state and to  analyze the critical behavior.

Recently, Wei et al. \cite{Wei:2021vdx} introduced topological methods through Duan's 
$\phi$-mapping theory \cite{Duan:1979ucg,Duan:1984ws}, assigning each critical point a topological charge $Q$ by treating it as a zero point of the topological current $j^\mu$ .
Their analysis uncovered two distinct classes: conventional critical points $Q=-1$ and novel critical points with
$Q=+1$, which is also called standard (vortex) and novel
(antivortex) critical point pair \cite{Ahmed:2022kyv}. 
Crucially, the first$-$order phase transition can emerge from the conventional critical points while the presence of the novel critical points cannot serve as an indicator of the first$-$order phase transition.
In 2023, Belhaj Ahmed et al. \cite{Ahmed:2022kyv} has revealed a third topological class: isolated critical points characterized by vanishing topological charge ($Q=0$). These occur in fine-tuned Lovelock gravity theories for hyperbolic black holes, exhibit nonstandard critical exponents, and are identified as topological phase transitions where vortex-antivortex pairs annihilate.

Although topological charges provide a robust classification, their deeper link to the dynamics of the system has not yet been examined systematically. 
To address this gap, we focus on quasinormal modes (QNMs) \cite{Wang:2005vs,Berti:2009kk,Konoplya:2011qq}—characteristic oscillations resulting from a black hole’s response to a perturbation.
Each QNM is characterized by a complex frequency $\omega = \omega_R + i\omega_I$, where $\omega_R$ governs oscillation frequency and $\omega_I$ determines the damping rate. 
%($\tau^{-1} = |\mathrm{Im}[\omega]|$).
In recent years researchers have folded QNMs into black-hole thermodynamics and shown that they act as dynamical fingerprints of phase transitions \cite{Gubser:2000mm,Jing:2008an,Berti:2008xu,He:2008im,Shen:2007xk,He:2010zb}.  
Numerical studies reveal a dramatic change in the slopes of quasinormal frequencies in small and large black holes near the first-order phase transition point where the Van der Waals-like thermodynamic phase transition occurs in RN-AdS black holes \cite{Liu:2014gvf}.
While near the critical isothermal phase transition point, the quasinormal frequencies for small and large black holes keep the same behavior as the black hole horizon increases.
These findings confirm that QNMs provide observable dynamical signatures of thermodynamic criticality.
We therefore employ QNMs to systematically investigate how topological charges manifest in black hole  dynamics.

To test whether critical points with distinct topological charges exhibit fundamentally
different dynamical responses or instead share universal behavior, we systematically study three representative black holes: charged RN-AdS as a prototype of conventional critical points ($Q=-1$), Born-Infeld-AdS (BI-AdS) \cite{Fernando:2003tz,Gunasekaran:2012dq} hosting both conventional ($Q=-1$) and novel ($Q=+1$) critical points and a quantum anomalous black hole \cite{Cai:2014jea,Hu:2024ldp} which provides the first known four-dimensional solution simultaneously hosting all three topological charges ($Q=-1,0$ and $+1$), while the lovelock gravity mentioned earlier \cite{Ahmed:2022kyv} can only be achieved in higher dimensions. 
We first identify the topological charges 
at their phase transition critical points. Then we compute the quasinormal  frequencies 
 near these critical points. 
 By comparing the results across topological classes, we will answer this question.

This paper is structured as follows. 
In Sec. II, we review the method for constructing the topological charge at critical points and use it to calculate the topological charge for different critical points of a quantum anomalous black hole. Meanwhile, we adopt the established results from Wei et al. \cite{Wei:2021vdx} for the RN-AdS and Born-Infeld-AdS systems. Section III details 
the quasinormal modes near the critical points of all three types of black holes.
By analyzing whether dynamical signatures correlate with topological classes or exhibit topology-independent universality, we resolve the core question of this work. Conclusions and discussions are presented in Sec. IV.

\section{Topological Charge of Critical Points}
In this section, we will review how to define the topological charge to characterize the topological properties of critical points in black holes and use this definition to systematically classify the critical points \cite{Wei:2021vdx}.

\subsection{General Framework of Topological Current Theory}
Following Duan’s $\phi$-mapping topological current theory \cite{Duan:1979ucg,Duan:1984ws}, 
the zero points of the vector field are very significance.
The topological current of the vector field $\phi$ is defined as
\begin{align}
    j^\mu=\frac{1}{2\pi}\epsilon^{\mu\nu\rho}\epsilon_{ab}\partial_\nu n^a\partial_\rho n^b,\quad\mu,\nu,\rho=0,1,2,
\end{align}
where $\partial_\nu=\frac{\partial}{\partial x^\nu}$, $x^\nu=(t,r,\theta)$ and the normalized vector
is defined as $n^a=\frac{\phi^a}{||\phi||}(a=1,2)$.
This current satisfies the conservation law 
$\partial_\mu j^\mu=0$.
By utilizing the Jacobi tensor \(\epsilon^{ab}J^{\mu}\left(\frac{\phi}{x}\right)=\epsilon^{\mu\nu\rho}\partial_\nu\phi^a\partial_\rho\phi^b\) and the two-dimensional Laplacian Green function \(\Delta _{\phi a} \ln \|\phi\| = 2\pi \delta(\phi)\), the topological current can be further written as:
\[
j^\mu = \delta^2(\phi) J^\mu \left( \frac{\phi}{x} \right)
\]
From this expression, it is zero wherever $\phi\neq0$, meaning it is non-vanishing only at the zero points of the vector field.
Finally, the corresponding topological charge in the given parameter region $\partial \sum$ can be calculated as 
\begin{align}
Q=\int_{\Sigma}j^0d^2x=\sum_{i=1}^N\beta_i\eta_i=\sum_{i=1}^Nw_i,
\end{align}
where  $\beta_{i}$ is the Hopf index of the $i$-th zero, $\eta_{i}=\pm1$ is its Brouwer degree and $w_i$
is the winding number.

Next, it naturally follows that if we can construct a vector field that links its zero points with the critical points in thermodynamics, we can study the topological properties of these critical points and classify them based on the value of the topological charge.
We now turn our attention to thermodynamics.

In black hole thermodynamics, according to the first law of thermodynamics
$dM=TdS+VdP+\sum_iY_idx^i$,
which leads to $T=(\frac{\partial M}{\partial S})_{P,x^i}$, and the Bekenstein-Hawking entropy formula $S=\frac{A}{4G}$,
Hawking temperature is typically expressed as a function of entropy, pressure and other parameters $x^i$, or equivalently as a function of pressure and thermodynamic volume.
In the following, $T(S,P,x^i)$ is used as an example.

Applying the first criticality condition $(\frac{\partial T}{\partial S})_{P,x^i}=0$, we solve for the pressure as $P=P(S,x^i)$.
Substituting this expression back gives 
$\tilde{T} (S,x^i)$ that depends only on $S$ and $x^i$.
Next, we introduce  the auxiliary angle $\theta$ and define
\begin{align}
    \Phi=\frac{1}{\sin \theta}\tilde{T}(S,x^i)\ .
\end{align}

Then we can construct a  vector field $\phi=(\phi^S,\phi^{\theta})=((\partial_S\Phi)_{\theta,x^i},(\partial_\theta\Phi)_{S,x^i})$.
The first advantage of the \(\theta\) term is that the direction of the introduced vector \(\phi\) is perpendicular to the horizontal lines at \(\theta = 0\) and \(\pi\), which can be treated as the boundaries in the parameter space. Another benefit is that the zero point of \(\phi\) is always located at \(\theta = \pi/2\). It is also straightforward to verify that the critical point coincides with the zero point of \(\phi\). This property is crucial as it enables the introduction of topology to study the black hole critical point.
\subsection{ TOPOLOGY OF BLACK HOLE SYSTEMS}
In this paper, we employ the foregoing topological framework to study the critical-point structure of three black holes. Our treatment of the RN–AdS and BI–AdS cases serves as a consistency check, as it reproduces known results \cite{Wei:2021vdx}.
In contrast, our analysis of the quantum-anomalous black hole is new and presented here as an original case study. 
Each black hole’s critical points belong to a different class which is why we chose them.

For RN-AdS black holes, after treating the cosmological constant as the pressure, the Hawking temperature is \cite{Kubiznak:2012wp}
\begin{align}
    T_{\rm RN}=\frac{2P\sqrt{S}}{\sqrt{\pi}}-\frac{\sqrt{\pi}q^2}{4S^{\frac{3}{2}}}+\frac{1}{4\sqrt{\pi}\sqrt{S}}\ ,
\end{align}
where $q$ is the charge of the black hole system and we set $q=1$ in the subsequent numerical calculations.
Specifically, its critical point is characterized by a topological charge of $Q=-1$, which corresponds to a conventional critical point.

For BI-AdS black holes, the Hawking temperature takes the following form \cite{Gunasekaran:2012dq}
\begin{align}
T_{\rm BI}=\frac{1}{4\sqrt{\pi^3S}}\left(2\tilde{b}^2S-2\sqrt{\tilde{b}^4S^2+\pi^2\tilde{b}^2q^2}+8\pi PS+\pi\right)\ ,
\end{align}
where $\tilde{b}$ represents the maximum electromagnetic field strength, and we take $q=1$ and $\tilde{b}=0.4$ in the following. 
Here we identify two critical points, which correspond to both conventional and novel types, with topological charges $Q=-1$ and $Q=1$, respectively.

Now we turn our attention to another quantum anomalous black hole, obtained by encoding quantum corrections through the conformal anomaly and feeding its back-reaction into Einstein’s equations. In curved spacetime the stress tensor develops a non-zero trace \cite{Duff:1993wm,Deser:1993yx},
\begin{align}
    \langle T^\mu{}_\mu \rangle = b\, I_4 - a\, E_4.
\end{align}
There are two different types of quantum trace anomalies. The Type A anomaly is associated with \(E_4\), describing quantum effects related to the curvature of the scalar and gravitational fields. The Type B anomaly is associated with \(I_4\), involving quantum effects related to the extension of fields and the structure of the gravitational field (such as electromagnetic fields). Here, we are keeping only the A-type contribution and writing $\alpha_c=8\pi a$ (we set $b=0$), the static spherically symmetric solution takes \cite{Cai:2009ua,Cai:2014jea}
\begin{align}
    ds^2=-f(r)dt^2+\frac{1}{f(r)}dr^2+r^2(d\theta^2+\sin^2\theta d\phi^2),
\end{align}
with
\begin{align}\label{metric}
    f(r)=1-\frac{r^2}{4\alpha_c}(1-\sqrt{1-8\alpha_c(\frac{2M}{r^3}-\frac{q^2}{r^4}-\frac{1}{l^2})})\ .
\end{align}
Here \(M\) and \(q\) are the mass and charge parameters, \(l\) is the AdS radius \((\Lambda=-3/l^{2})\), and \(\alpha_{c}\) measures the strength of the anomaly. In the limit \(\alpha_{c}\to 0\), the metric smoothly reduces to the RN--AdS form.

The Hawking temperature can be expressed through the horizon radius \(r_h\) \cite{Hu:2024ldp}:
\begin{equation}
T=\frac{3r_h^{4}/l^{2}+r_h^{2}-q^{2}+2\alpha_{c}}{4\pi r_h^{3}-16\pi\alpha_{c}r_h}.
\end{equation}

In the extended phase space, 
after identifying cosmological constant as the thermodynamic pressure 
$P=-\frac{\Lambda}{8\pi}=\frac{3}{8\pi l^{2}}$,  one can easily obtain the equation of state in the
extended phase space
\begin{align}
T=\frac{2 \alpha _c+8 \pi  P r_h^4-q^2+r_h^2}{4 \pi  r_h \left(r_h^2-4 \alpha _c\right)}.
\end{align}
which shares analogous behavior with the four-dimensional charged Gauss–Bonnet black hole in AdS space \cite{Fernandes:2020rpa}. 
Nevertheless, the physical interpretations of $\alpha_c$ and $q$ are distinct.

After eliminating 
$P$ as described earlier,
The corresponding thermodynamic function reads
\begin{align}
    \Phi=\frac{1}{\sin \theta}\left(\frac{4 \alpha _c-2 q^2+r_h^2}{2 \pi  r_h^3-24 \pi  r_h \alpha _c}\right)
\end{align}
The components of the vector field $\phi$ are
\begin{align}
    \phi^{r_h}&=-\frac{\csc\theta\left(-6q^2r_h^2+r_h^4+24\left(q^2+r_h^2-2\alpha_c\right)\alpha_c\right)}{2\pi\left(r_h^3-12r_h\alpha_c\right)^2},\\
    \phi^{\theta}&=\frac{\cot \theta \csc \theta (2 q^2 - r_h^2 - 
   4 \alpha_c)}{
2 \pi r_h^3 - 24 \pi r_h \alpha_c}.
\end{align}
Then one can easily obtain the normalized vector field  through $n=(\frac{\phi^{r_{h}}}{||\phi||},\frac{\phi^\theta}{||\phi||})$. 

From the metric Eq.(\ref{metric}), imposing both a non-negative central charge and the existence of a critical point, one can restrict the coupling constant to $0\leq \alpha_c \leq q^2/8$ \cite{Hu:2024ldp}.
In the limit $\alpha_c$ approaches zero, there is only one critical point and can return to the results in RN-AdS black hole, which has been discussed earlier. When $\alpha_c=q^2/8$, there is also only one critical point. However, this case is very different from the $\alpha \rightarrow0$ case, which will be discussed in detail in the following part.

%Thus we will only focus on the following two other cases.

\begin{figure}[h]
  \centering
  \includegraphics[width=0.6\textwidth]{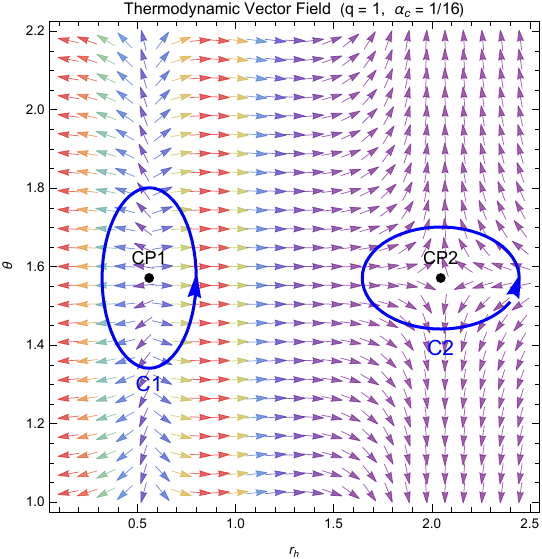}
\captionsetup{font=small}
  \caption{
The rainbow-colored arrows represent the vector field $n$ on a portion of the $\theta-r_h$ plane with $\alpha_c=1/16$ and $q=1$.
The critical points $CP_{1}$ and
$CP_{2}$ located at $(r_h-\theta)=(0.560, \pi/2)$ and $(2.046, \pi/2)$  are marked
with black dots, and they are enclosed with the blue contour $C_1$ and $C_2$,  respectively. The topological charge on the $CP_{1}$ (antivortex) is 1, the one on the right (vortex) is -1.
 }\label{fig1}
\end{figure}

For $0<\alpha_c<q^2/8$, there are two critical points,  the behavior of the normalized vector field $n$ is  exhibited in Fig.\ref{fig1} with $\alpha_c=1/16$ and $q=1$ as an example.
In order to calculate the topological charge, we shall construct two contours $C_1$ and $C_2$, which are parametrized by $\theta \in(0,2\pi)$ as
\begin{align}
\begin{cases}
r=a\cos\vartheta+r_0, \\
\theta=b\sin\vartheta+\frac{\pi}{2}. & 
\end{cases}
\end{align}
We choose $(a,b,r_0)$=$(0.240,0.230,0.560)$ for $C_1$, and $(0.400,0.130,2.046)$ for $C_2$.
A new measure that quantifies how the vector field bends along the chosen contour is introduced
\begin{align}
\Omega(\vartheta)=\int_0^\vartheta\epsilon_{ab}n^a\partial_\vartheta n^bd\vartheta,
\end{align}
where $\epsilon_{ab}$ denotes the two-dimensional Levi-Civita symbol.
Then the topological charge must be $Q=\frac{1}{2\pi}\Omega(2\pi).$
Therefore the topological charge $Q_{CP_{1}}=1$ and $Q_{CP_{2}}=-1$  for the critical points $CP_{1}$ and $CP_{2}$.

\begin{figure}[h]
  \centering
\includegraphics[width=1.05\textwidth]{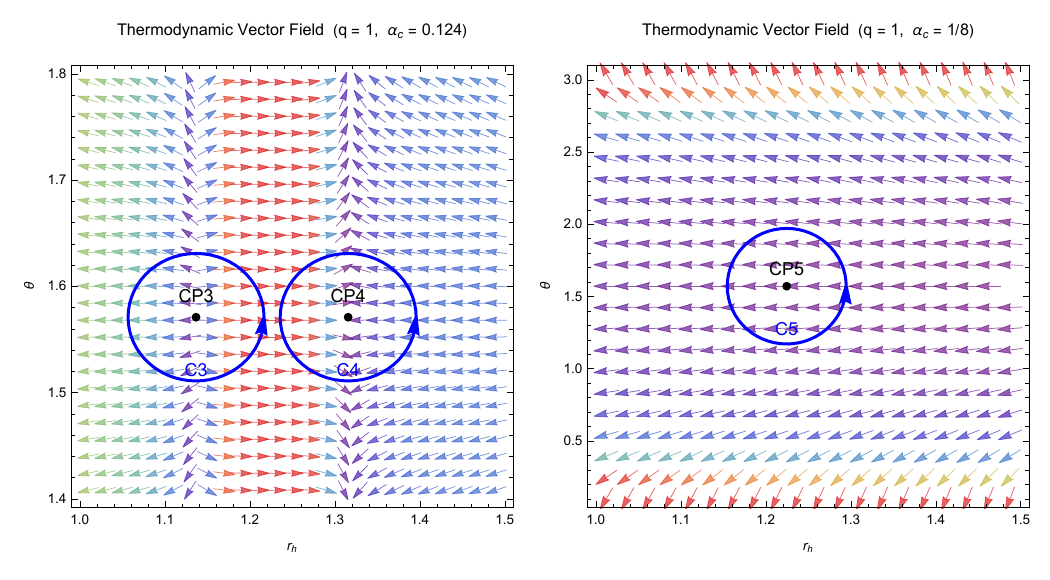}
\captionsetup{font=small}
  \caption{
The rainbow-colored arrows represent the vector field $n$ on a portion of the $\theta-r_h$ plane  for two values of $\alpha_c$.
Left: setting $\alpha_c=0.124$, there are two critical points positioned very close to each other.
The left-hand point (antivortex) carries a positive topological charge, whereas the right-hand point (vortex) carries a negative one.
Right: setting $\alpha_c=1/8$,
 there are no fixed points of $n^a$, and the $CP_5$ which is also called ICP and marked by a black dot, possesses zero topological charge, as confirmed by integrating along the blue contour shown in the figure.
}\label{fig2}
\end{figure}

For $\alpha_c=q^2/8$, 
the two critical points merge into a single isolated critical point, leaving only one critical point,
which makes $\alpha_c=q^2/8$ a very special case.
Apart from the observation of critical point merging, the system exhibits several other intriguing thermodynamic properties, including the violation of scaling laws, which have been discussed in detail in Ref.\cite{Hu:2024ldp}.
We respectively plotted two cases in Fig.\ref{fig2} for $\alpha_c=0.124$ (left)  and $\alpha_c=1/8$ (right), to demonstrate in detail how this merger occurs.
The left panel shows the same situation discussed above, while the right one exhibits a different zero point—an isolated critical point.
We find that the isolated critical point does not  obey
($\partial T/ \partial r_h)_{P}=0$,  instead, this point is the limit point approached $\alpha_c=1/8$. 
It is endowed as expected with zero topological charge,
\begin{align}
    Q(CP_5)=0.
\end{align}
As the coupling $\alpha_c$ rises to its critical value, the existing vortex–antivortex pair gradually moves toward each other until it finally disappears. If we then lower $\alpha_c$ back past this point, a new vortex–antivortex pair forms.

As we mentioned earlier, this isolated critical point has also been found in the phase diagram of hyperbolic black holes in odd-order Lovelock theories and can be understood as the onset of the standard (vortex) and novel (antivortex) critical point pair creation \cite{Ahmed:2022kyv}.
Compared to previous studies, the quantum anomalous black hole in present work is four-dimensional, and this is the first time a critical point with topological charge 0 has been found in lower-dimensional spacetimes, which is worth mentioning.

Thus far, we have identified three distinct critical points in this black-hole system: a conventional point with $Q=-1$, a novel one  with $Q=1$ and an isolated critical one with $Q=0$.

\section{The behavior of quasinormal frequencies at the critical point with different topological classification}
After investigating the topological thermodynamic properties of the critical points, we now turn our focus to the dynamic properties near these points, specifically the QNMs \cite{Fu:2022cul,Fu:2023drp,Zhu:2024wic}.

\subsection{Massless Perturbations in Black Hole Spacetime}
We first examine a massless scalar perturbation in the spacetime of each black hole.
Writing the perturbation as $\Psi(r,t)=\psi(r)e^{-i\omega t}$, the radial function $\psi(r)$ satisfies the Klein–Gordon equation
\begin{align}\label{KG}
    \psi^{\prime\prime}(r)+\left[\frac{f^{\prime}(r)}{f(r)}+\frac{2}{r}\right]\psi^{\prime}(r)+\frac{\omega^2\psi(r)}{f(r)^2}=0,
\end{align}
where $\omega$ denotes the oscillation frequency of the perturbation.

Near the event horizon $r_H$, we impose the ingoing boundary condition so that the total perturbation behaves as $\psi(r)\sim(r-r_H)^{-i\omega/(4\pi T)}$.
To factor out this oscillatory phase and eliminate the inconvenient $\omega^2/f(r)^2$ term, we write $\psi(r)=\exp\left[-i\int^{r}\frac{\omega}{f(\tilde{r})}d\tilde{r}\right]\tilde{\phi}(r)$, where the exponential reproduces the ingoing behaviour and 
$\tilde{\phi}(r)$ is regular at the horizon.
Substituting this ansatz into Eq.(\ref{KG}), we cancel the common exponential factor and arrive at
\begin{align}\label{3.2}
    \tilde{\phi}^{\prime\prime}(r)+\tilde{\phi}^{\prime}(r){\left[\frac{f^{\prime}(r)}{f(r)}+\frac{2}{r}-2i\frac{\omega}{f(r)}\right]}-\tilde{\phi}\frac{2i\omega}{rf(r)}=0.
\end{align}
Because the phase factor has already enforced the ingoing condition, we can simply normalize $\tilde{\phi}(r_H)=1$, while at the AdS boundary $r\to\infty$, we impose $\tilde{\phi}(\infty)=0$.
Since Eq.(\ref{3.2}) holds for any static, spherically symmetric metric, it can be directly applied to all three black-hole solutions considered here.
It  is then solved numerically by a pseudospectral method, which numerical computations are performed with the QNMspectral package \cite{Jansen:2017oag}, to obtain the quasinormal frequencies.

\subsection{ The Behavior of Quasinormal Frequencies at the Critical Point}
In the following study, the critical parameters of the isothermal phase transition have been determined through numerical analysis. Tables 1–4 list the quasinormal mode  of massless scalar perturbations for black holes of different sizes, evaluated precisely at the critical temperature 
$T=T_c$. Within each table, the rows above the horizontal divider correspond to small black holes, whereas the rows below refer to large black holes.

For RN-AdS Black Holes,
We consider the case with the critical radius \(r_H = \sqrt{6}q\) and here \(q = 1\) as we mentioned earlier, for which the topological charge is \(Q = -1\). The results are reported in Table 1.

\begin{table}[h]
\centering
\begin{tabular}{|c|c|c|c|}
\hline
\( r_H \) & \( 
\omega_R
 \) & \( \omega_I \) \\
\hline
2.43724 & 0.377441638 & -0.184081512 \\
2.44214 & 0.377505914 & -0.184432473 \\
2.44459 & 0.377538114 & -0.18460802 \\
2.44704 & 0.377570352 & -0.18473608 \\
\hline
2.45194 & 0.377634937 & -0.185134903 \\
2.45439 & 0.377667283 & -0.185310606 \\
2.45684 & 0.377699661 & -0.185486344 \\
2.46174 & 0.37776451  & -0.185837916 \\
\hline
\end{tabular}
\caption{Quasinormal frequencies of small and large RN-AdS black holes at the critical isothermal point $T=T_c=\frac{\sqrt{6}}{18\pi }$ with $Q=-1$. Values above the horizontal line are for small black holes; those below are for large black holes.}
\label{tab1}
\end{table}

Turning to BI-AdS black holes,
we examine the case where $\tilde{b}=0.4$ and  \(q = 1\), two critical points can be identified, with the corresponding critical radii being $r_H=1.11422$ ($Q=1$) and $r_H=2.08089$ ($Q=-1$). The results are shown in Table 2.
\begin{table}[h]
\centering
\begin{tabular}{|c|c|c|}
\hline
\( r_H \) & \( \omega_R \) & \( \omega_I \) \\
\hline
1.07044 & 0.295997 & -0.04905 \\
1.08399 & 0.296074 & -0.04955 \\
1.09372 & 0.296121 & -0.04991 \\
1.10308 & 0.296163 & -0.05026 \\
\hline
1.12535 & 0.296252 & -0.05108 \\
1.13869 & 0.296304 & -0.05157 \\
1.14825 & 0.296343 & -0.05193 \\
1.16018 & 0.296395 & -0.05238 \\
\hline
\end{tabular}
\hspace{1cm}
\begin{tabular}{|c|c|c|}
\hline
\( r_H \) & \( \omega_R \) & \( \omega_I \) \\
\hline
1.99626 & 0.397419 & -0.1762 \\
2.01826 & 0.39776 & -0.17796 \\
2.03967 & 0.398095 & -0.17967 \\
2.05988 & 0.398415 & -0.1813 \\
\hline
2.1017  & 0.399083 & -0.18467 \\
2.12271 & 0.39942  & -0.18637 \\
2.14172 & 0.399725 & -0.18791 \\
2.16333 & 0.400071 & -0.18966 \\
\hline
\end{tabular}
\caption{Quasinormal frequencies of small and large BI-AdS black holes at the critical isothermal point $T=T_c=0.04049$ with $Q=1$ (left) and $T=T_c=0.04589$ with $Q=-1$ (right). Values above the horizontal line are for small black holes; those below are for large black holes.}

  \label{tab2}
\end{table}

In the case of the quantum anomalous black hole, the situation becomes more complex. We will consider two scenarios: 
$\alpha_c=1/16$ and $\alpha_c=1/8$. The results for the first scenario will be presented in Table. 3, where two critical points are observed. 
The corresponding critical radii are $r_H=0.55992$ ($Q=1$) and $r_H=2.04609$ ($Q=-1$).
The results for the second scenario will be shown in Table. 4, where only one isolated critical point remains with $r_H=1.22474$ ($Q=0$).
\begin{table}[h!]
\centering
\begin{tabular}{cc}
\begin{tabular}{|c|c|c|}
\hline
\( r_H \) & \( \omega_R \) & \( \omega_I \) \\
\hline
0.53753 & 3.378508 & -2.17848 \\
0.54393 & 3.493132 & -2.21325 \\
0.54893 & 3.575244 & -2.24053 \\
0.55432 & 3.657809 & -2.27002 \\
\hline
0.56552 & 3.813576 & -2.33151 \\
0.57091 & 3.882329 & -2.36117 \\
0.57711 & 3.957276 & -2.39531 \\
0.58292 & 4.023963 & -2.4273 \\
\hline
\end{tabular}
\hspace{1cm}
\begin{tabular}{|c|c|c|}
\hline
\( r_H \) & \( \omega_R \) & \( \omega_I \) \\
\hline
1.96425 & 0.465541 & -0.22529 \\
1.98471 & 0.466083 & -0.22747 \\
2.00517 & 0.466634 & -0.22967 \\
2.02563 & 0.46719 & -0.23188 \\
\hline
2.06655 & 0.468312 & -0.23633 \\
2.08701 & 0.468873 & -0.23857 \\
2.10747 & 0.469434 & -0.2408 \\
2.12793 & 0.469992 & -0.24303 \\
\hline
\end{tabular}
\end{tabular}
\caption{Quasinormal frequencies of small and large quantum anomalous black holes at the critical isothermal point $T=T_c=\frac{\sqrt{9-2 \sqrt{15}} \left(\sqrt{15}+4\right)}{3 \pi }$ with $Q=1$ (left) and $T=T_c=\frac{3-\frac{\sqrt{15}}{2}}{3 \sqrt{\frac{\sqrt{15}}{2}+\frac{9}{4}} \pi }$ with $Q=-1$ (right), with $\alpha=1/16$. Values above the horizontal line are for small black holes; those below are for large black holes.}

  \label{tab3}
\end{table}

\begin{table}[h!]
\centering
\begin{tabular}{|c|c|c|}
\hline
\( r_H \) & \( \text{Re} \, \omega \) & \( \text{Im} \, \omega \) \\
\hline
1.17535 & 0.923196 & -0.49106 \\
1.18879 & 0.927254 & -0.49647 \\
1.19944 & 0.930427 & -0.5008 \\
1.21249 & 0.934261 & -0.50616 \\
\hline
1.23798 & 0.941572 & -0.51676 \\
1.24943 & 0.944777 & -0.52156 \\
1.26309 & 0.948533 & -0.52732 \\
1.27292 & 0.951192 & -0.53147 \\
\hline
\end{tabular}
\caption{Quasinormal frequencies of small and large quantum anomalous black holes at the critical isothermal point $T=T_c=1/(\sqrt{6} \pi)$ with $Q=0$, with $\alpha=1/8$. Values above the horizontal line are for small black holes; those below are for large black holes.}

  \label{tab4}
\end{table}
As we all known, the real part represents the oscillation frequency of the perturbation. The imaginary part  determines the damping rate of the oscillation, and the corresponding damping time \(\tau\) is defined as \(\tau = 1/|\omega_I|\), which represents the time required for the oscillation to decay to \(1/e\) of its initial value. The negative sign of the imaginary part directly reflects the system's stability: when \(\omega_I < 0\), the system is stable.

From these tables, we can discern a clear feature: regardless of the value of $Q$, the quasinormal modes of black holes near critical points display a universal monotonic trend. 
Specifically, both the real part \(\omega_R\) and the imaginary part  \( |\omega_I|\) increase as the horizon radius \(r_H\) increases. 
This behavior suggests that 
the oscillation period and decay time are both decreasing, which means the system responds more quickly and recovers stability faster.
At the critical points of different topological charges, both the real and imaginary parts increase as the event horizon radius grows. 

Based on initial observations, it is challenging to distinguish the differences between the topological charges solely from the quasinormal modes. However, rather than concluding at this point, we conducted a more detailed analysis to further explore the similarities and differences between them.

Therefore, we calculated the specific response and sensitivity of the quasinormal modes. The response of the QNMs can be quantitatively represented by the derivatives of the oscillation frequency and damping rate with respect to the event horizon radius, typically using the following formulas:
$d\omega_R/d r_H$ and $ d|\omega_I|/d r_H$.
To obtain these derivatives, we performed a least-squares fitting of the data and derived the slopes.
On the other hand, 
we define elasticity parameters
\[
\text{Elasticity}(\omega_R) = \frac{\overline{r_H}/d r_H}{\overline{\omega_R}/d \omega_R}, \quad \text{Elasticity}= \frac{\overline{r_H}/d r_H}{\overline{\omega_I}/d \omega_I}.
\]
to quantify the sensitivity, which also means the relative variation of the quasinormal frequency with respect to the event horizon radius of black holes.
%Similar sensitivity analyses can be found in Ref. \cite{Konoplya:2022iyn}.

After classifying all the results by topological charge, we organized the data in Table~\ref{tab5} and draw the following conclusions.
First, we focus on $d\omega_R/dr_H$. Even with different black holes, the critical point for \(Q = -1\) exhibits similar behavior, with values very close to each other, all around the \(10^{-2}\) order of magnitude, which are \(7.2\times10^{-2}\),\ \(8.1\times10^{-2}\) and \(10.9\times10^{-2}\). In contrast, when examining the results for \(Q = 0\) and \(Q = 1\), none of them fall within this range, the order of magnitude instead being \(10^{-3}\), \(10^{-1}\), and \(10^1\), respectively. 
Furthermore, for \(Q = 1\), the two different black holes show significant differences, with no similar behavior observed.
When we observe $d|\omega_I|/dr_H$ and elasticity, the same situation also occurs.

We conclude that this is not a coincidence and \(Q = -1\) cases may belong to the same category. 
On one hand, \(Q = -1\) shows similarity, while on the other hand, \(Q = 1\) does not exhibit such similarity.
This is exactly the thermodynamic topological imprint we aim to identify in the dynamics.

\begin{table}[htbp]
\centering
\label{tab:QNM_all}
\small
\begin{tabular}{|c|c|c|c|}
\hline
Topology \& BH
& $d\omega_R/dr_H$
& $d|\omega_I|/dr_H$
& Elasticity $(\omega_R, \omega_I)$

\\ \hline
$Q=-1$ \; (RN-AdS )
& $1.3\times 10^{-2}$
& $7.2\times 10^{-2}$
& 9\%, 95\%

\\
$Q=-1$ \; (BI-AdS)
& $1.5\times 10^{-2}$
& $8.1\times 10^{-2}$
& 8.3\%, 91.6\%

\\
$Q=-1$ \; (QA)
& $2.7\times 10^{-2}$
& $10.9\times 10^{-2}$
& 11.9\%, 94.8\%

\\ \hline
$Q=0$ \; (QA)
& $2.9\times 10^{-1}$ 
& $4.1\times 10^{-1}$ 
& 37.5\%,\; 99.3\% 

\\ \hline
$Q=+1$ \; (BI-AdS)
& $4.3\times 10^{-3}$
& $3.7\times 10^{-2}$
& 1.6\%,81.5\%

\\
$Q=+1$ \; (QA)
& $1.4\times 10^{1}$
& $0.5\times 10^{1}$
& 2.12,\; 1.33

\\ \hline
\end{tabular}
\caption{Summary of linear scalings of QNMs near criticality for $Q=-1,0,+1$. We report the slopes with respect to $r_H$ and elasticities at the sample mean.}
\label{tab5}
\end{table}

\section{Conclusion and Discussions}
In this study, we explored the connection between the topological classification of critical points in black hole thermodynamics and quasinormal modes. Using Duan’s $\phi$-mapping theory, we first classified the critical points of the quantum anomalous black holes.
It is worth emphasizing that this is also the first time an isolated critical point has been discovered in a four-dimensional black hole.

After performing a topological classification of the critical points for three representative black holes, we then calculated the quasinormal modes in their vicinity. We found that, in terms of the trends of the real and imaginary parts with respect to the event horizon radius, they appear to be identical, both increasing as the radius grows. However, upon further analysis, we observed certain differences in their rate of change and elasticity: when the topological charge is -1, they exhibit a high degree of similarity and can be categorized into a unified class, while significant differences are observed when \(Q=1\) and \(Q=0\). 

There is still some work to be done in the future.
First, although our current conclusion seems to be based on a limited sample size and appears to be case-by-case, we plan to calculate the critical points of more black holes with a topological charge of -1. This will help us determine if they follow the patterns established in this paper. Moreover, we aim to provide a purely analytical, or at least semi-analytical, explanation for why they appear to belong to the same category.
Second, the impact of topological charge on other types of perturbations, such as electromagnetic or gravitational waves, should be further investigated. 
Third, studying QNMs in higher-dimensional black holes and in the presence of more complex quantum effects may provide deeper insights into the role of topology in black hole physics.

Overall, this study suggests a connection between thermodynamic topology and the dynamics of black holes, offering a new perspective on black hole thermodynamics.

\section*{Acknowledgement}

We are grateful to Dr.Guoyang Fu and Dr. Ligang Zhu for helpful discussions. Ya-Peng Hu is supported by
National Natural
Science Foundation of China (NSFC) under grant Nos.12175105.


\begin{thebibliography}{99}
%\cite{Natsuume:2014sfa}
\bibitem{Natsuume:2014sfa}
M.~Natsuume,
%``AdS/CFT Duality User Guide,''
Lect. Notes Phys. \textbf{903} (2015), pp.1-294
doi:10.1007/978-4-431-55441-7
[arXiv:1409.3575 [hep-th]].
%213 citations counted in INSPIRE as of 29 Sep 2025

%\cite{Harlow:2014yka}
\bibitem{Harlow:2014yka}
D.~Harlow,
%``Jerusalem Lectures on Black Holes and Quantum Information,''
Rev. Mod. Phys. \textbf{88} (2016), 015002
doi:10.1103/RevModPhys.88.015002
[arXiv:1409.1231 [hep-th]].
%493 citations counted in INSPIRE as of 29 Sep 2025

%\cite{Hawking:1982dh}
\bibitem{Hawking:1982dh}
S.~W.~Hawking and D.~N.~Page,
%``Thermodynamics of Black Holes in anti-De Sitter Space,''
Commun. Math. Phys. \textbf{87} (1983), 577
doi:10.1007/BF01208266
%2894 citations counted in INSPIRE as of 29 Sep 2025

%\cite{Dolan:2011xt}
\bibitem{Dolan:2011xt}
B.~P.~Dolan,
%``Pressure and volume in the first law of black hole thermodynamics,''
Class. Quant. Grav. \textbf{28} (2011), 235017
doi:10.1088/0264-9381/28/23/235017
[arXiv:1106.6260 [gr-qc]].
%717 citations counted in INSPIRE as of 29 Sep 2025

%\cite{Kubiznak:2012wp}
\bibitem{Kubiznak:2012wp}
D.~Kubiznak and R.~B.~Mann,
%``P-V criticality of charged AdS black holes,''
JHEP \textbf{07} (2012), 033
doi:10.1007/JHEP07(2012)033
[arXiv:1205.0559 [hep-th]].
%1260 citations counted in INSPIRE as of 29 Sep 2025

%\cite{Wei:2021vdx}
\bibitem{Wei:2021vdx}
S.~W.~Wei and Y.~X.~Liu,
%``Topology of black hole thermodynamics,''
Phys. Rev. D \textbf{105} (2022) no.10, 104003
doi:10.1103/PhysRevD.105.104003
[arXiv:2112.01706 [gr-qc]].
%155 citations counted in INSPIRE as of 29 Sep 2025

%\cite{Duan:1979ucg}
\bibitem{Duan:1979ucg}
Y.~S.~Duan and M.~L.~Ge,
%``SU(2) Gauge Theory and Electrodynamics with N Magnetic Monopoles,''
Sci. Sin. \textbf{9} (1979) no.11, 1072
doi:10.1142/9789813237278{\_}0001
%84 citations counted in INSPIRE as of 29 Sep 2025

%\cite{Duan:1984ws}
\bibitem{Duan:1984ws}
Y.~Duan,
%``THE STRUCTURE OF THE TOPOLOGICAL CURRENT,''
SLAC-PUB-3301.
%57 citations counted in INSPIRE as of 29 Sep 2025




%\cite{Ahmed:2022kyv}
\bibitem{Ahmed:2022kyv}
M.~B.~Ahmed, D.~Kubiznak and R.~B.~Mann,
%``Vortex-antivortex pair creation in black hole thermodynamics,''
Phys. Rev. D \textbf{107} (2023) no.4, 046013
doi:10.1103/PhysRevD.107.046013
[arXiv:2207.02147 [hep-th]].
%51 citations counted in INSPIRE as of 29 Sep 2025


%\cite{Wang:2005vs}
\bibitem{Wang:2005vs}
B.~Wang,
%``Perturbations around black holes,''
Braz. J. Phys. \textbf{35} (2005), 1029-1037
doi:10.1590/S0103-97332005000700002
[arXiv:gr-qc/0511133 [gr-qc]].
%47 citations counted in INSPIRE as of 29 Sep 2025


%\cite{Berti:2009kk}
\bibitem{Berti:2009kk}
E.~Berti, V.~Cardoso and A.~O.~Starinets,
%``Quasinormal modes of black holes and black branes,''
Class. Quant. Grav. \textbf{26} (2009), 163001
doi:10.1088/0264-9381/26/16/163001
[arXiv:0905.2975 [gr-qc]].
%2162 citations counted in INSPIRE as of 29 Sep 2025

%\cite{Konoplya:2011qq}
\bibitem{Konoplya:2011qq}
R.~A.~Konoplya and A.~Zhidenko,
%``Quasinormal modes of black holes: From astrophysics to string theory,''
Rev. Mod. Phys. \textbf{83} (2011), 793-836
doi:10.1103/RevModPhys.83.793
[arXiv:1102.4014 [gr-qc]].
%1438 citations counted in INSPIRE as of 29 Sep 2025

%\cite{Gubser:2000mm}
\bibitem{Gubser:2000mm}
S.~S.~Gubser and I.~Mitra,
%``The Evolution of unstable black holes in anti-de Sitter space,''
JHEP \textbf{08} (2001), 018
doi:10.1088/1126-6708/2001/08/018
[arXiv:hep-th/0011127 [hep-th]].
%352 citations counted in INSPIRE as of 29 Sep 2025

%\cite{Jing:2008an}
\bibitem{Jing:2008an}
J.~Jing and Q.~Pan,
%``Quasinormal modes and second order thermodynamic phase transition for Reissner-Nordstrom black hole,''
Phys. Lett. B \textbf{660} (2008), 13-18
doi:10.1016/j.physletb.2007.11.039
[arXiv:0802.0043 [gr-qc]].
%77 citations counted in INSPIRE as of 29 Sep 2025

%\cite{Berti:2008xu}
\bibitem{Berti:2008xu}
E.~Berti and V.~Cardoso,
%``Quasinormal modes and thermodynamic phase transitions,''
Phys. Rev. D \textbf{77} (2008), 087501
doi:10.1103/PhysRevD.77.087501
[arXiv:0802.1889 [hep-th]].
%32 citations counted in INSPIRE as of 29 Sep 2025

%\cite{He:2008im}
\bibitem{He:2008im}
X.~He, B.~Wang, S.~Chen, R.~G.~Cai and C.~Y.~Lin,
%``Quasinormal modes in the background of charged Kaluza-Klein black hole with squashed horizons,''
Phys. Lett. B \textbf{665} (2008), 392-400
doi:10.1016/j.physletb.2008.06.038
[arXiv:0802.2449 [hep-th]].
%45 citations counted in INSPIRE as of 29 Sep 2025

%\cite{Shen:2007xk}
\bibitem{Shen:2007xk}
J.~Shen, B.~Wang, C.~Y.~Lin, R.~G.~Cai and R.~K.~Su,
%``The phase transition and the Quasi-Normal Modes of black Holes,''
JHEP \textbf{07} (2007), 037
doi:10.1088/1126-6708/2007/07/037
[arXiv:hep-th/0703102 [hep-th]].
%51 citations counted in INSPIRE as of 29 Sep 2025

%\cite{He:2010zb}
\bibitem{He:2010zb}
X.~He, B.~Wang, R.~G.~Cai and C.~Y.~Lin,
%``Signature of the black hole phase transition in quasinormal modes,''
Phys. Lett. B \textbf{688} (2010), 230-236
doi:10.1016/j.physletb.2010.04.006
[arXiv:1002.2679 [hep-th]].
%37 citations counted in INSPIRE as of 29 Sep 2025

%\cite{Liu:2014gvf}
\bibitem{Liu:2014gvf}
Y.~Liu, D.~C.~Zou and B.~Wang,
%``Signature of the Van der Waals like small-large charged AdS black hole phase transition in quasinormal modes,''
JHEP \textbf{09} (2014), 179
doi:10.1007/JHEP09(2014)179
[arXiv:1405.2644 [hep-th]].
%129 citations counted in INSPIRE as of 29 Sep 2025

%\cite{Fernando:2003tz}
\bibitem{Fernando:2003tz}
S.~Fernando and D.~Krug,
%``Charged black hole solutions in Einstein-Born-Infeld gravity with a cosmological constant,''
Gen. Rel. Grav. \textbf{35} (2003), 129-137
doi:10.1023/A:1021315214180
[arXiv:hep-th/0306120 [hep-th]].
%202 citations counted in INSPIRE as of 07 Oct 2025

%\cite{Gunasekaran:2012dq}
\bibitem{Gunasekaran:2012dq}
S.~Gunasekaran, R.~B.~Mann and D.~Kubiznak,
%``Extended phase space thermodynamics for charged and rotating black holes and Born-Infeld vacuum polarization,''
JHEP \textbf{11} (2012), 110
doi:10.1007/JHEP11(2012)110
[arXiv:1208.6251 [hep-th]].
%676 citations counted in INSPIRE as of 07 Oct 2025

%\cite{Cai:2014jea}
\bibitem{Cai:2014jea}
R.~G.~Cai,
%``Thermodynamics of Conformal Anomaly Corrected Black Holes in AdS Space,''
Phys. Lett. B \textbf{733} (2014), 183-189
doi:10.1016/j.physletb.2014.04.044
[arXiv:1405.1246 [hep-th]].
%78 citations counted in INSPIRE as of 07 Oct 2025

%\cite{Hu:2024ldp}
\bibitem{Hu:2024ldp}
Y.~P.~Hu, Y.~S.~An, G.~Y.~Sun, W.~L.~You, D.~N.~Shi, H.~Zhang, X.~Chen and R.~G.~Cai,
%``Quantum anomaly triggers the violation of scaling laws in gravitational system,''
[arXiv:2410.23783 [gr-qc]].
%11 citations counted in INSPIRE as of 07 Oct 2025

%\cite{Duff:1993wm}
\bibitem{Duff:1993wm}
M.~J.~Duff,
%``Twenty years of the Weyl anomaly,''
Class. Quant. Grav. \textbf{11} (1994), 1387-1404
doi:10.1088/0264-9381/11/6/004
[arXiv:hep-th/9308075 [hep-th]].
%635 citations counted in INSPIRE as of 07 Oct 2025

%\cite{Deser:1993yx}
\bibitem{Deser:1993yx}
S.~Deser and A.~Schwimmer,
%``Geometric classification of conformal anomalies in arbitrary dimensions,''
Phys. Lett. B \textbf{309} (1993), 279-284
doi:10.1016/0370-2693(93)90934-A
[arXiv:hep-th/9302047 [hep-th]].
%513 citations counted in INSPIRE as of 07 Oct 2025

%\cite{Cai:2009ua}
\bibitem{Cai:2009ua}
R.~G.~Cai, L.~M.~Cao and N.~Ohta,
%``Black Holes in Gravity with Conformal Anomaly and Logarithmic Term in Black Hole Entropy,''
JHEP \textbf{04} (2010), 082
doi:10.1007/JHEP04(2010)082
[arXiv:0911.4379 [hep-th]].
%181 citations counted in INSPIRE as of 07 Oct 2025

%\cite{Fernandes:2020rpa}
\bibitem{Fernandes:2020rpa}
P.~G.~S.~Fernandes,
%``Charged black holes in AdS spaces in 4D Einstein Gauss-Bonnet gravity,''
Phys. Lett. B \textbf{805} (2020), 135468
doi:10.1016/j.physletb.2020.135468
[arXiv:2003.05491 [gr-qc]].
%263 citations counted in INSPIRE as of 21 Oct 2025

%\cite{Fu:2022cul}
\bibitem{Fu:2022cul}
G.~Fu, D.~Zhang, P.~Liu, X.~M.~Kuang, Q.~Pan and J.~P.~Wu,
%``Quasinormal modes and Hawking radiation of a charged Weyl black hole,''
Phys. Rev. D \textbf{107} (2023) no.4, 044049
doi:10.1103/PhysRevD.107.044049
[arXiv:2207.12927 [gr-qc]].
%23 citations counted in INSPIRE as of 07 Oct 2025

%\cite{Fu:2023drp}
\bibitem{Fu:2023drp}
G.~Fu, D.~Zhang, P.~Liu, X.~M.~Kuang and J.~P.~Wu,
%``Peculiar properties in quasinormal spectra from loop quantum gravity effect,''
Phys. Rev. D \textbf{109} (2024) no.2, 026010
doi:10.1103/PhysRevD.109.026010
[arXiv:2301.08421 [gr-qc]].
%36 citations counted in INSPIRE as of 07 Oct 2025

%\cite{Zhu:2024wic}
\bibitem{Zhu:2024wic}
L.~G.~Zhu, G.~Fu, S.~Li, D.~Zhang and J.~P.~Wu,
%``Quasinormal modes of a charged loop quantum black hole,''
Phys. Rev. D \textbf{111} (2025) no.10, 10
doi:10.1103/PhysRevD.111.104008
[arXiv:2410.00543 [gr-qc]].
%12 citations counted in INSPIRE as of 07 Oct 2025

%\cite{Jansen:2017oag}
\bibitem{Jansen:2017oag}
A.~Jansen,
%``Overdamped modes in Schwarzschild-de Sitter and a Mathematica package for the numerical computation of quasinormal modes,''
Eur. Phys. J. Plus \textbf{132} (2017) no.12, 546
doi:10.1140/epjp/i2017-11825-9
[arXiv:1709.09178 [gr-qc]].
%214 citations counted in INSPIRE as of 07 Oct 2025













\end{thebibliography}
\end{document}